# A Ku-BAND NOVEL MICROMACHINED BANDPASS FILTER WITH TWO TRANSMISSION ZEROS

*Zhang Yong[1], Zhu Jian[1], Yu Yuanwei[1], Chen Chen[1,2], Jia Shi Xing[1]*
[1] Nanjing Electronic Devices Institute, Nanjing, 210016, PRC
[2] Dept. of Physics, Nanjing University, 210093, PRC
TEL: 86-25-86858306, email: j-zhu@public1.ptt.js.cn

**ABSTRACT**

This paper presents a micromachined bandpass filter with miniature size that has relatively outstanding performance. A silicon-based eight-order microstrip bandpass filter is fabricated and measured. A novel design method of the interdigital filter that can create two transmission zeros is described. The location of the transmission zeros can be shifted arbitrarily in the stopband. By adjusting the zero location properly, the filter provides much better skirt rejection and lower insertion loss than a conventional microstrip interdigital filter. To reduce the chip size, through-silicon-substrate-via-hole is used. Good experimental results are obtained.

## 1. INTRODUCTION

Microwave filters are widely used in communication applications to reject spurious signals and to separate different channels in a multichannel communication system. So the characteristics of compact size, high selectivity, and low insertion loss for microwave filters are highly required. To meet these objectives, some key techniques are used in this paper, such as tapped-line interdigital structure, micromachined via-hole and implementation of transmission zeros at finite frequency.

In tapped-line filters design, the first and the end coupling sections are eliminated. Therefore, filters with tapped-line feed can offer space and cost saving advantages. Furthermore, the realizable bandwidth could be much greater [1].

Through-silicon-substrate-via-hole is realized by MEMS technique. It is used for the grounding of the microwave resonators in the filter and therefore makes the filter much more compact [2].

To build in prescribed transmission zeros can remarkably improve the selectivity of the filters. Various methods have been introduced for the implementation of the transmission zeros at finite frequencies. Using cross couplings between nonadjacent resonators is the most common way [3]. But it usually suffers more complex structures in design. To avoid cross coupling structures, open-loop and hairpin transmission-line resonator filters using tapped-line feed topology to create transmission zeros have also been reported [4], [5]. However, their size is usually not as compact as the one with interdigital structure.

In this paper, a Ku band micromachined tapped-line interdigital bandpass filter with two transmission zeros is presented. It is easy to realize in the physical structure. Each transmission zero can be freely placed at desired frequency to achieve demand attenuation in the stopband. The locations of the transmission zeros can be calculated. All the grounded sections of the filter are realrized by micromachined via-hole. Finally, an eight-order bandpass filter is fabricated and measured. The measured performance shows good agreement with the desin theory. The chip size is only 9.6mm×4mm.

## 2. ANALYSES AND DESIGN OF Ku-BAND MICROMACHINED BANDPASS FILTERS

### 2.1 Micromachined interdigital filters with via-hole

The basic structure of the filter is using interdigital microstrip quarter-wave lines with short-circuited line at the ends. The filter specification is 8 poles, 0.2dB Chebyshev ripple, and 15% fractional bandwidth. Table. 1. shows the normalized element values (from $g_0$ to $g_9$) for Chebyshev response. [6]

Table. 1. Normalized element values.

| $g_0$ | $g_1$ | $g_2$ | $g_3$ | $g_4$ |
|---|---|---|---|---|
| 1 | 1.3804 | 1.3875 | 2.2963 | 1.5217 |
| $g_5$ | $g_6$ | $g_7$ | $g_8$ | $g_9$ |
| 2.3413 | 1.4925 | 2.1349 | 0.8972 | 1.5386 |

The coupling coefficient (k) is related as

$$k_{j,j+1} = \frac{w}{\sqrt{g_j g_{j+1}}} \qquad (1)$$

where

$$w = \frac{f_2 - f_1}{f_0}$$

$f_2$ and $f_1$ are the high and low resonant frequencies [7]. The mutual coupling matrix (M) can be calculated from (1). Then, the physical size ($w_0$, $w_1$, $w_2$… $w_8$, $w_9$, $s_{01}$, $s_{12}$, $s_{23}$…$s_{78}$, $s_{89}$) of filter can be calculated [7].

$$M = \begin{bmatrix} 0 & 0.144 & 0 & 0 & 0 & 0 & 0 & 0 \\ 0.144 & 0 & 0.112 & 0 & 0 & 0 & 0 & 0 \\ 0 & 0.112 & 0 & 0.107 & 0 & 0 & 0 & 0 \\ 0 & 0 & 0.107 & 0 & 0.106 & 0 & 0 & 0 \\ 0 & 0 & 0 & 0.106 & 0 & 0.107 & 0 & 0 \\ 0 & 0 & 0 & 0 & 0.107 & 0 & 0.112 & 0 \\ 0 & 0 & 0 & 0 & 0 & 0.112 & 0 & 0.144 \\ 0 & 0 & 0 & 0 & 0 & 0 & 0.144 & 0 \end{bmatrix}$$

Micromachined via-hole is used for the grounding at the end of the quarter wave microstrip transmission line. The hole shape and size can also affect the filter performance. It should be chosen carefully. The discussion of the via-hole has been reported explicitly in [2]. After optimized by electromagnetic (EM) simulation, the final structure of the filter is obtained. Fig. 1. shows the sketch of the filter. The length of the resonator (L) is 1730 µm.

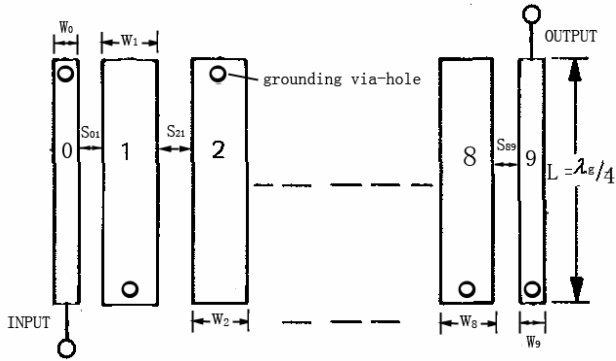

Fig. 1. Micromachined interdigital bandpass filter with via-hole.

**2.2 Micromachined tapped-line interdigital filters**

After the physical size of the filter is determined, the first and end coupling lines are then changed into tapped-line structure.

Fig. 2. shows the micromachined interdigital filter with tapped-line feed.

The tap point $l/L$ can be calculated from

$$\frac{Q_e}{Z_0/Z_{0I}} = \frac{\pi}{4\sin^2(\pi l/2L)} \quad (2)$$

$$Q_e = \frac{f_0}{\Delta f_{\pm 90°}} \quad (3)$$

where $Z_0$ is the characteristic impedance of the input and output line, $Z_{0I}$ is the characteristic impedance of the resonators, $Q_e$ is the external $Q$ of the resonator, $f_0$ is the resonator frequency, and $\Delta f_{\pm 90°}$ is the bandwidth about the resonant frequency over which the phase varies from $-90°$ to $+90°$ [8]. After optimization, the tapping position of $l/L = 950\mu m/1730\mu m = 0.55$ is obtained. And the size of the resonators ($w_1$, $w_2$… $w_8$, $s_{12}$, $s_{23}$…$s_{78}$) is also optimized. Fig.3. shows the simulated and measured results. The filter works in Ku band. It has a fractional 3-dB bandwidth of 20%. The insertion loss is 2.1dB at 13.2GHz, and the return loss is greater than 15dB within the pass band. The attenuation is greater than 25.5dB extend to 11 and 13 GHz and beyond.

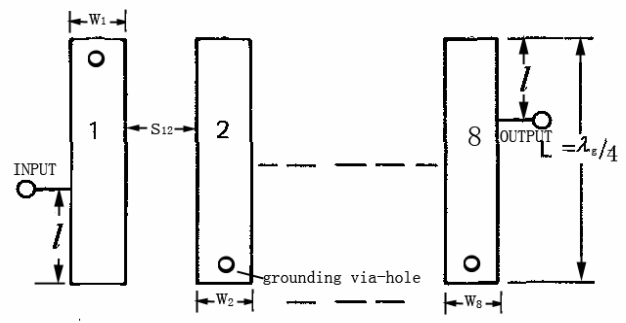

Fig. 2. Micromachined tapped-line interdigital bandpass filter.

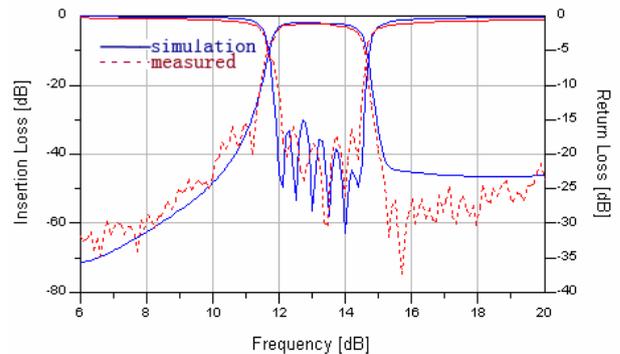

Fig. 3. Simulated and measured results of the micromachined tapped-line interdigital filter.

The measured results show a relatively good performance and make a good agreement with the simulated ones. However, the attenuation on lower side is still need to be improved. This problem can be solved by the realization of the transmission zero.

**2.3 Micromachined tapped-line interdigital filters with two transmission zeros**

As open-loop and hairpin filters are, the micromachined interdigital filter with two transmission zeros in this paper is also on the base of the tapped-line feed topology. However, the first and end sections are skillfully changed. As shown in Fig. 2., each resonator element of the traditional tapped-line interdigital filters is a quarter-wave length long at midband frequency and is short-circuited at one end and open-circuited at the other end. As a result, there can't be two transmission zeros close to the passband. Thus a new structure is proposed by changing the first and end resonators into transmission lines with both ends short-circuited. That is, Both the first and the end section have two grounded points.

The length of the first and end transmission lines can be longer than the other resonators of the filter, and also the width may be changed a lot. The location of transmission zeros can be calculated from

$$f = \frac{c}{2l\sqrt{\varepsilon_{re}}} \quad (4)$$

where $l$ is the length from one grounded point of the first or the end transmission lines to the corresponding tap point. $\varepsilon_{re}$ is the effective permittivity.

To achieve high attenuation and low insertion loss, the transmission zeros desired should be within the frequency of 10 GHz and 11 GHz on the lower side, 15 GHz and 16 GHz on the upper side. So the length is calculated within 3860 μm and 4250 μm for the transmission zero on the lower side, 2830 μm and 2650 μm for the transmission zero on the upper side. However, the location of the transmission zeros can also affect the coupling condition of the first and the end section. It may cause the overcoupled situation [6]. Thus there will be a hump in the passband. To get proper coupling condition, the transmission zeros can be tuned in a very narrow frequency range and it just affect the property of the filter very slightly. The critically coupled condition is given by

$$K = \frac{1}{Q_e} + \frac{1}{Q_u} \quad (5)$$

where $K$ is the coupling coefficient, $Q_e$ is the external $Q$, and $Q_u$ is the unloaded $Q$ of the resonator [7]. After calculation, the final structure of the filter is obtained and then electromagnetic (EM) simulation is done. The comparison of simulated results of two types of filters is shown in Fig. 4. A is the results of the filter described in Chapter 2.2 and B is that of the filter with two transmission zeros. It shows that the improvement of the out-of-band attenuation is remarkble. In A, the attenuation at 10.6 GHz is 36.8 dB. While in B, it is 48.3dB at the same frequency.

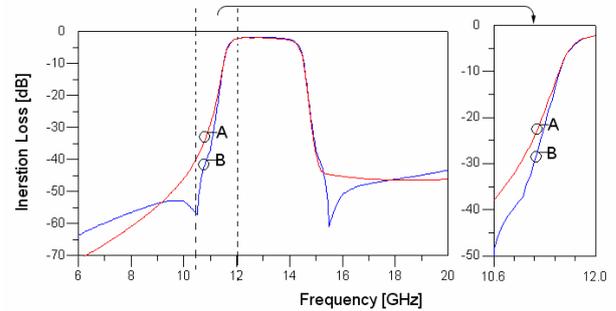

Fig. 4. Simulated results of two types of filters A and B.

To achieve more compact size, matched microstrip corners are introduced. That is, the first and the end lines are designed into folded structure. That affects the location of the transmission zeros very slightly.

## 3. EXPERIMENTAL RESULTS OF FABRICATED FILTERS

The filter is fabricated on the 4″ high resistivity silicon substrate. Inductively coupled plasma (ICP) process is used for the fabrication of the silicon via-hole. From the experiment results shown in Fig. 5, the fractional 3-dB bandwidth remains unchanged 20% and the return loss is greater than 15dB. While the attenuation is greater than 37dB extend to 11 and 13 GHz and beyond. It is nearly improved by 12dB. Furthermore, the insertion loss that is 1.9dB is also improved by 0.2dB. The experiment results make a good agreement with the simulated ones. The chip size is only 9.6mm×4mm. The photograph of the fiter is shown in Fig. 6.

## 4. CONCLUSIONS

A micromachined tapped-line interdigital filter is designed, fabricated and measured. The explicit design procedure is described. All the grounded sections of the filter are realized by micromachined via-hole. The filter has two transmission zeros due to its novel structure. Measured results have shown the excellent performance and advantages of this type of filters. And the filter size is much smaller than the conventional one.

This product is protected by PRC patent NO. 200510094411.2.

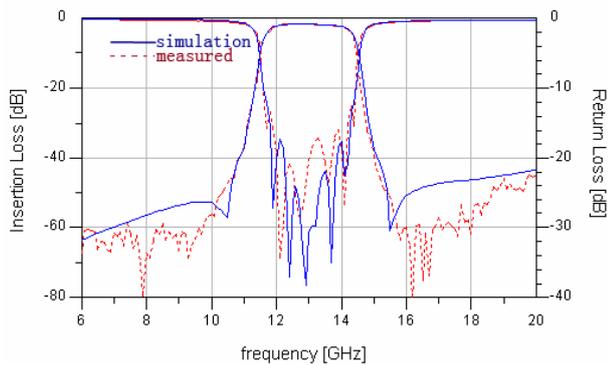

Fig.5. Measured and simulated results of micromachined tapped-line interdigital filter with two transmission zeros.

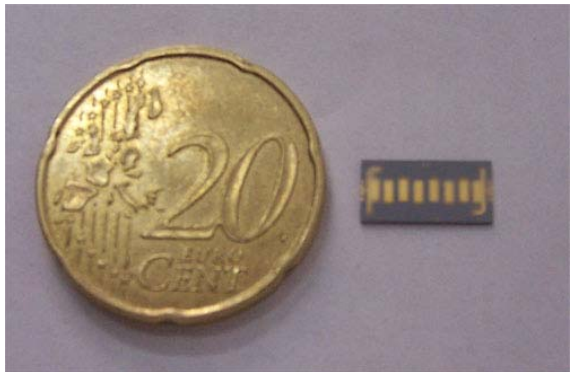

Fig. 6. Photograph of a fabricated filter

## 5. REFERENCE

[1] Kai Chang, *Handbook of RF/Microwave Components And Engineering*. Hoboken, New Jersey. John Wiley & Sons, 2003, ch3.

[2] Jian Zhu, "Micromachined Silicon Via-hole and Interdigital Bandpass Filters," in DTIP of MEMS/MOEMS 2005, pp.86-89.

[3] R. Levy, "Filters with single transmission zeros at real or imaginary frequencies," IEEE Trans. Microwave Theory Tech, vol. AP-24, pp.172-181, Apr. 1976.

[4] Lung-Hwa Hsueh, "Tunable Microstrip Bandpass Filters With Two Transmission Zeros," IEEE Tans.Microwave Theory Tech, vol 51,pp.520-524, Feb. 2003

[5] C.-M.Tsai, S.-Y. Lee, and C.-C. Tsai,"Hairpin filters with tunable transmission zeros," in IEEE MTT-S Int. Microwave Symp. Dig., 2001, pp.2175-2178.

[6] Giovanni Alessio, "Interdigital Design Forms Low-cost Bnadpass Filters," MICROWAVES&RF, NO. 77-85, SEP 1997.

[7] G.L. Matthaci, L.Young, and E. M. T. Jones. *Micowave Filters, Impedance-Maching Networks, and Coupling Structures*. New York. McGraw-Hill, 1980, ch11.

[8] J.S Wong,"Microstrip tapped-line filter design," IEEE Trans. Microwave Theory Tech, vol 27,pp.45-50, Jan. 1979.